\documentclass{article}

\usepackage{PRIMEarxiv}

\usepackage[utf8]{inputenc} 
\usepackage[T1]{fontenc}    
\usepackage{hyperref}       
\usepackage{url}            
\usepackage{booktabs}       
\usepackage{amsfonts}       
\usepackage{nicefrac}       
\usepackage{microtype}      
\usepackage{lipsum}
\usepackage{fancyhdr}       
\usepackage{graphicx}       
\graphicspath{{media/}}     
\usepackage[flushleft]{threeparttable}
\usepackage{xcolor,colortbl}
\usepackage{amsmath,amsfonts,amsthm}
\usepackage{tabularx,rotating}
\usepackage{caption}
\usepackage{epsfig}
\usepackage{subcaption}
\usepackage{calc}
\usepackage{verbatim}
\usepackage{amssymb}
\usepackage{amstext}
\usepackage{amsmath}
\usepackage{amsthm}
\usepackage{multicol}
\usepackage{multirow}
\usepackage{algorithm}
\usepackage{longtable}
\usepackage{float}
\usepackage{array}
\usepackage{graphicx}

\title{Data Querying with Ciphertext Policy Attribute Based Encryption
}

\author{
  {Maryam Almarwani, Boris Konev and   Alexei Lisitsa}\\
 Department of Computer Science\\
 University of Liverpool\\
 Liverpool, U.K\\
  \texttt{\{M.almarwani, Konev,  A.Lisitsa\}@liverpool.ac.uk} }

\begin{document}
\maketitle

\begin{abstract}
Data encryption limits the power and efficiency of queries. Direct processing of encrypted data should ideally be possible to avoid the need for data decryption, processing, and re-encryption. It is vital to keep the data searchable and sortable. That is, some information is intentionally leaked. This intentional leakage technology is known as "querying over encrypted data schemes", which offer confidentiality as well as querying over encrypted data, but it is not meant to provide flexible access control.
This paper suggests the use of Ciphertext Policy Attributes Based Encryption (CP-ABE) to address three security requirements, namely: confidentiality, queries over encrypted data, and flexible access control. By combining flexible access control and data confidentiality, CP-ABE can authenticate who can access data and possess the secret key. Thus, this paper identifies how much data leakage there is in order to figure out what kinds of operations are allowed when data is encrypted by CP-ABE.
\end{abstract}


\section{{Introduction}}

\label{sec:introduction}
Outsourced storage is often used by data owners, such as organisations or individuals, to store their data and process it via an outsourced service provider. Such data can be vulnerable to external attackers and the outsourced service provider (a curious administrator). Therefore, sensitive data should be encrypted before it is sent to outsourced storage. There are two principal encryption deployment methods on outsourced servers. The first of which is server-side encryption, wherein data is encrypted after receipt but prior to writing and storage in outsourced storage. The second method is client-side encryption, whereby data is encrypted on the client side before it is transmitted to the outsourced storage. Traditional encryption algorithms such as symmetric and asymmetric cryptography are currently used for server and client-side encryption.Querying over encrypted data schemes comprise symmetric and asymmetric encryption that permit a limited number of operations over encrypted data while ensuring confidentiality.

In this paper, satisfying three security requirements \cite{ferretti2013access}, which are confidentiality, queries over encrypted data, and flexible access control, has been considered at the data level.
In general, traditional encryption, such as querying over encrypted data schemes , is not meant to provide flexible access control.
Flexible access control grants access to data parts in accordance with the policy of the data owner. This policy can be altered as required without impacting the encrypted data. When multiple users need to access encrypted data, they must either store the keys on the outsourced service provider's server-side encryption or exchange the keys between parties (i.e., users) for client-side encryption. The keys' storage does not prevent an attack because they may be used in an honest-but-curious manner or clearly misused.
All encrypted data becomes vulnerable if the secret key is leaked (i.e., exposed by an attacker). As a consequence, in principle, the key may be shared with multiple users, but this comes with a considerable risk.
This is in contrast to asymmetric encryption, often known as public key encryption, wherein data owners use different keys to encrypt and decrypt data. Parties who use asymmetric key encryption can share the encryption key with each other, but keep the decryption key secret.
Asymmetric encryption is used when two or more parties exchange data by encrypting it through one party’s public key and transferring it to the other party, who then decrypts it using the other party’s private key. To address queries through encryption sharing data, a secret key must be shared. If this secret key is revealed, anybody who possess it has the ability to decrypt all of the data belonging to that party.
As a result, traditional asymmetric encryption is unsuitable for data that a data owner shares with multiple users.
Therefore, a mechanism is required to enforce access control on decryption operations, such as Ciphertext Policy Attributes Based Encryption (CP-ABE)  \cite{bethencourt2007ciphertext}. 
This type of encryption ensures that only the owner of the secret key that matches the access policy is able to perform decryption. By combining flexible access control with data confidentiality, CP-ABE authorizates who can access data and possession of secret keys. The data owner’s access policy consists of a set of data access conditions based on attributes. CP-ABE has been used to exchange data,
see e.g.
\cite{jahid2012piratte}. However, CP-ABE has never been assessed in relation to the querying of encrypted data \cite{inbook},\cite{Release},\cite{almarwani2021release},\cite{release3}.
Hence, it is essential to ascertain the level of data leakage in order to identify the supported operations involved in the querying of CP-ABE encrypted data. For this reason, this paper demonstrates how CP-ABE can meet three security requirements, wherein the efficiency of CP-ABE   is compared to AES.
The following points are the focus of this paper:
\begin{itemize}
\item We illustrate the use and efficiency of CP-ABE. We present the results of an evaluation of the executed queries on data encrypted with CP-ABE. 
\item In addition, this paper assesses 
the use of \emph{ Ciphertext Policy Attribute-Based Encryption} on  data in order to 
address the needs for flexible access control, confidentiality, and querying encrypted data at the same time.  
\item Furthermore, we compare the security and performance of CP-ABE against that of AES, which is often used on databases.
\end{itemize}
The rest of the paper is divided into five additional sections, the first of which is the background  including Ciphertext Policy Attribute-Based Encryption (Section \hyperref [sec:ch4A.2]{2}). Database encryption methods are outlined in Section \hyperref[sec:ch4A.3]{ 3}. The access control mechanisms are discussed in Database in section \hyperref[sec:ch4A.4]{4}, which comprises a comparison of CP-ABE efficiency and AES, which includes an evaluation of performance, security, and functionality.  A summary is presented in Section \hyperref[sec:ch4A.6]{5}.

\section{Background and Related Work}

 \label{sec:ch4A.2}
     \subsection{Querying over encrypted data schemes}
     \label{sec:ch3.3}
 Encryption is a commonly used approach for protecting sensitive data. While data encryption improves security, it can impose limitations on query execution. Therefore, direct data processing should be enabled to prevent data decryption, processing, and re-encryption through some disclosed information. This approach is called "Querying over encrypted data".
It  is to protect data in ciphertexts while  allowing some of the features that allow ciphertext computation.
 The following are the most commonly used querying over encrypted data schemes:

\begin{enumerate}
\item	{\bf Random (RND) \cite{popa2011cryptdb} :}  
two identical values are (very likely)  encrypted into different ciphertexts using a random value.  It can be achieved, e.g by applications  
of Blowfish-CBC or AES-CBC algorithms, using a random initialization vector (IV).
Using RND 
does not expose any information about encrypted values and does not allow for any computation on them.
The level of protection supported by RND encryption can be characterized 
as Indistinguishability under Chosen plain-text Attack  ({\bf{IND-CPA}})\cite{bellare2004introduction}. 
\item	{\bf Deterministic (DET) \cite{popa2011cryptdb}:} this 
type of encryption ensures 
that two identical values are encrypted into two identical ciphertexts.  It can be ashieved, e.g. by using 
AES  encryption in CBC mode with fixed IV (e.g. IV = 0). 
Hence,it results in leaking the information on equal  values, and hence, one is able to conduct equality check  operations on the encrypted data, including equality predicate, Count, and Group. Because of that 
it grants reduced security compared to RND 
which can be characterized 
as INDistinguishability under the Distinct Chosen-Plaintext Attack {\bf{(IND-DCPA)} \cite{boldyreva2011order}}.

\item	{\bf Order-Preserving Encryption (OPE)  \cite{agrawal2004order}:} 
Order-Preserving encryption, such as implemented e.g. in 
\cite{boldyreva2011order} enables comparison predicates (e.g. Order by, Min, Max)  on encrypted values.  
It 
offers less security as compared to DET, which can be characterized as  INDistinguishability under Ordered Chosen-Plain-text Attack{\bf{ (IND-OCPA)} \cite{boldyreva2011order}}. 
\item	{\bf {Searchable Encryption  \cite{song2000practical}}:} 
this kind of encryption 
enables one to search through a collection of encrypted data without 
using 
decryption. 
For example, 
Song et al.  \cite{song2000practical} propose the schema 
for 
single keyword search  over encrypted data  that 
supports the execution of the Like operation in queries. 
Provided level of security protection is 
{\bf{IND-CPA}}
\cite{bellare2004introduction}.
\item	{\bf Homomorphic encryption (HOM)  \cite{arasu2014querying}:} 
this type  of encryption  
enables users to perform computations on encrypted data without the need to decrypt it.
Partially homomorphic encryption (PHE)  can support particular operations, such as summation or multiplication.  
Pailier's algorithm \cite{paillier1999public} 
supports summation operation. 
Fully homomorphic encryption (FHE) on the other hand can support both addition and multiplication \cite{gentry2009fully}, but at the cost of efficiency.

HOM emcryption generally provides  level of security similar to that of to RND,   
that is  IND-CPA.

\end{enumerate}

Previous secure database research (  {Hacigumus et.al} \cite{hacigumucs2002executing},  {{CryptDB }} \cite{popa2011cryptdb}, 
{{Monomi}} \cite{tu2013processing},
{{BlindSeer}} \cite{pappas2014blind},
{{L-EncDB}} \cite{li2015encdb},
 {{DBmask}} \cite{sarfraz2015dbmask}, {{ZeroDB}} 
\cite{egorov2016zerodb}\cite{mitterer2018experimental},
{{CryptMDB}} \cite{xu2017cryptmdb}, 
  {{CloudDBGuard}} \cite{wiese2020clouddbguard},
{{Arx}} \cite{poddar2016arx}, 
 {{CryptGraphDB}}  \cite{aburawi2018querying}, and  D-SDDB  \cite{icissp19}) have all based on one or more querying over encrypted data schemes.
 
\subsection{Ciphertext-Policy Attribute-Based Encryption}
Attributes-Based Encryption (ABE) \cite{bethencourt2007ciphertext} 
 is an advanced public key cryptographic method wherein encryption and decryption are based on attributes. ABE's restrictions on data access as per data owner imposed attributes are defined by the access policy. ABE can be classified into Key-Policy Attribute-Based Encryption (KP-ABE)  and Ciphertext-Policy Attribute-Based Encryption (CP-ABE).
 The ciphertext is associated with a set of attributes in KP-ABE, whereas the secret key is associated with the access policy. An access policy established over the attributes of the encrypting party is encoded into a user's attribute secret key in KP-ABE, and a ciphertext is generated with regard to an attribute list. An attribute secret key can only be used to decrypt ciphertext if the ciphertext attribute list matches the access policy that is  associated with the user's attribute secret key. The issue with the KP-ABE technique is that the encrypting party  does not have control over who can decrypt the encrypted data. It can only choose descriptive attributes for the data and therefore should trust a secret key party. KP-ABE is more suited to passive users who can't be  controlled  to decrypt by the encrypting party  and active ciphertext, which is pushed to users \cite{motkpabe}. An example is television programme broadcasting, where the encrypting party  demands fast encryption but has no control over the number or identity of receivers.
In contrast, a user's secret key is associated with an attribute list in CP-ABE, and ciphertext defines an access policy defined over the encrypting party's attribute. A user can decrypt a ciphertext if and only if the user's attribute list in the secret key matches the access policy of the ciphertext.  Thus, the encrypting party controls who can decrypt data.
CP-ABE is suited for active users who can be controlled to decrypt by the encrypting party and passive ciphertext, for which users actively seek ciphertext \cite{motkpabe}. Cloud data is an example where the encrypting party can accept slower encryption but select who the expected receivers are.
In concept, CP-ABE is more similar to standard access control models such as Role-Based Access Control (RBAC) \cite{bethencourt2007ciphertext}. Thus, CP-ABE is more natural to employ instead of KP-ABE for enforcing access control on encrypted data.
\begin{itemize}
    \item {\bf{CP-ABE Model:}} 
    In existing CP-ABE schemes, there are several types of access policies, such as monotone access formulas consisting of AND, OR (Boolean), threshold gates, or non-monotone access formulas consisting of Linear Secret Sharing Scheme (LSSS).
    This paper employs a simplified version of the BSW \cite{bethencourt2007ciphertext} scheme, which formed the foundation for the initial CP-ABE model proposal and other existing CP-ABE models.
    BWS supports the monotone access formula, which cannot express negative constraints, via a Boolean formula (i.e. the access tree ($T$)). 
\begin{itemize}
\item {\bf{Access tree $T$:}}
 Private keys are identified with a set $S$ of descriptive attributes in the BWS model  \cite{bethencourt2007ciphertext}. A party wishing to encrypt a message will provide an access policy (an access tree $T$) that private keys must satisfy in order to decrypt it. 
Each non-leaf node represents a threshold gate, which is defined by its children and a threshold value. If $num_x$ is the number of children of node $x$ and $k_x$ is its threshold value where $0 <k_x \le num_x$ . An attribute and a threshold value $k_x = 1$ are assigned to each tree leaf node $x$. 
The satisfaction of an access tree $T$ with the root $r$ by a set of attributes $\gamma$ is defined as follows:

Let $T_x$ be the sub-tree of $T$ that is rooted at node $x$. 
Define the function $T_x(\gamma)$ (``$\gamma$ satisfies subtree $T_{x}$'')  recursively as follows. If $x$ is non-leaf node, then $T_{x}(\gamma) = 1$ if $|\{x'| \mbox{x' is a child of x} \land T_{x'}(\gamma) = 1| \ge k_{x}$.

If $x$ is a leaf node then  $T_x (\gamma) = 1$ 
if and only if $att (x)$  $ \in \gamma$.

\end{itemize}
    \item {\bf{CP-ABE Algorithms:}}
 The four basic algorithms in the BWS scheme
 are listed below.
    \begin{enumerate}
\item {\bf{Setup: }} This selects a $G_0$ bilinear group, which are groups with an efficient bilinear map \cite{bethencourt2007ciphertext} and a generator $g$ of prime order $p$. It will then choose two random exponents $\alpha$, $\beta$ $\in$ $Z_p$. This algorithm includes only the implicit security parameter input ($g$,$p$,$\alpha$, $\beta$) while outputting the public parameters $pk$ as well as a master key $mk$.
\item {\bf{Key Generation ($mk, S$):}} This algorithm takes 
the master key $mk$  and
a set of attributes $S$ that will describe  the private key $sk$ 
as input while outputting a private key $sk$. 
The algorithm first selects a random $r$ $\in$ $Z_p$, followed
by a random $r_j$ $\in$   $Z_p$  for each attribute $j$  $\in$ $S$. The secret key is then computed.
\item {\bf{Encrypt ($pk, T, M$):}} The encryption algorithm encrypts a message $M$ using the tree access structure $T$. 
\item {\bf{Decrypt($pk, CT, sk$):}} This algorithm takes the public parameters pk, a ciphertext $CT$ which contains an access policy $T$ 
, and a private key $sk$ 
related to a set $S$ of attributes as input. When the set $S$ of attributes fulfils the access structure $A$, the algorithm can decrypt the ciphertext and return a message $M$.
\end{enumerate}

\end{itemize}
\begin{itemize}
    \item {\bf{How it Works \cite{touati2014c}}}: 
    \begin{figure}[ht]
\centerline{\includegraphics[width=.4\textwidth,height=.2\paperwidth]{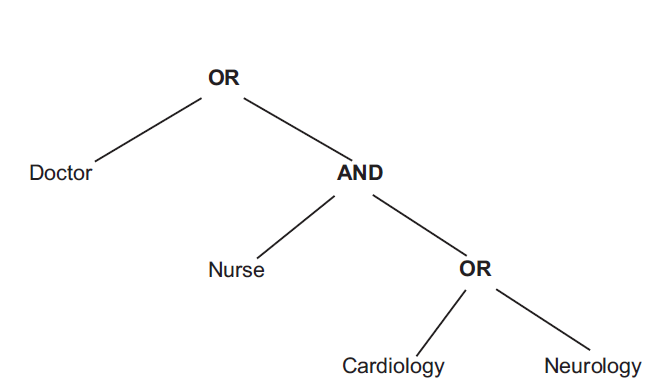}}
\caption{An example of an access tree  \cite{touati2014c}}
\label{fig:tree}
\end{figure}
Consider a system in which a group of users $U = \{u_1, u_2, \ldots, u_n\}$ have access to shared data based on an access policy. The access policy is defined by an access structure such as the one shown in Figure ~\ref{fig:tree}.  
According to their role in the system, each user will be assigned a subset of the attributes $D_i = \{d_1, d_2, \ldots , d_x\}$.
An Attribute Authority $AA$ manages the set of attributes 
utilized in the $A = \{a_1, a_2, \ldots , a_k\}$ (where $\forall i \in \{1, \ldots , n\},D_i \subseteq  A$).
The AA generates a Public Key $pk$ and distributes it to all system entities during the bootstrap phase, then generates a Master Key $mk$ that AA keeps  secret {\bf{(Algorithm 1)}}. The AA then assigns a secret Key sk to each user in the system in accordance with their attributes {\bf{(Algorithm 2)}}. Once this phase is complete, each system user is provided with a secret key based on their attribute set, which also provided the public key and the complete set of attributes.
When a user wishes to share data with other users in the system, the user builds up the corresponding access tree and encrypts the data using the encryption algorithm {\bf{ (Algorithm 3)}}. The system verifies 
an access policy in which the user does not know the identities of the users who are supposed to read the data a priori.
The ciphertext is then sent to a remote server. Users that wish to access the data must have the set S of attributes in their private key satisfy the access policy in a ciphertext CT. 
If this does not occur, it becomes impossible to decrypt the data {\bf{(Algorithm 4)}}.
\item {\bf{Security:}} CP-ABE provides security against {\bf{collusion attacks}} \cite{bethencourt2007ciphertext}. Collusion attempts will fail because the private keys of each user are generated randomly. 
A random number $r$ in Key Generation {\bf{(Algorithm 2)}} is selected for each user and embedded in all of that user's private keys by AA. As a result, if a user attempts to utilise private keys from other users, the user will not be utilising the same random $r$ through the decryption process, and hence it will fail.
\end{itemize}
For more detailed information, see \cite{zhang2020attribute} for a comprehensive discussion of CP-ABE and its classification.  Enhanced CP-ABE incorporates all the cryptographically functional features of basic CP-ABE (BSW scheme \cite{bethencourt2007ciphertext}), in addition to supplementary features.  

\section{Encryption on Databases}

 \label{sec:ch4A.3}
This section provides an overview of the encryption algorithms that are often used to prevent data leakage, as well as the encryption methods that are used to deploy encryption algorithms.

 \subsection{Encryption Algorithms}
 \label{sec:ch4A.3.1}
Symmetric and asymmetric encryption are the two main types of encryption algorithms. Symmetric encryption is ideal for encrypting huge amounts of data quickly. Even so, since it utilises the same key for encryption and decryption, it does not enable identity verification. Asymmetric encryption, on the other hand, employs a key pair to ensure that the data is only read by the intended receiver. This verification, however, slows down the encryption process.
Database engines ranking  \cite{Ranking}  shows the algorithms employed for the most common database engines, where the encryption algorithms currently in use can be symmetric (i.e.  DES\cite{feistel1973cryptography}, Blowfish\cite{schneier1993description} ,RC2\cite{schneier1994blowfish}, AES\cite{daemen1999aes}) or/ and  asymmetric (i.e. RSA\cite{rivest1978textordfemininea}, ElGamal encryption\cite{elgamal1985public}). In addition, the AES encryption algorithm
is commonly used for both relational and non-relational databases.

 \subsection{Encryption deployment methods}
  \label{sec:ch4A.3.2}
A database encryption procedure can be 
deployed in a number of ways that are all the way from the database engine to the application layer. 
The three most common database encryption deployment methods are Application Programming Interface (API), Plug-In, and Transparent Data Encryption (TDE) \cite{EncrMethods}.
 \begin{itemize}
     \item {\bf{The API Method :}} The application layer is where the API method encryption occurs.
     To implement the encryption, the engineer must use a supplied code with a modification function in order to change the web server’s defined method. The API methods operate in conjunction with any database without adding any additional load to the database management system (DBMS).
     As a consequence, all queries that reference encrypted data
     in the application must be modified explicitly.

    \item {\bf{The Plug-In Method :}} The Plug-In method requires the DBMS to have an encryption package attachment (i.e., User Defined Functions (UDFs)). The DBMS encryption package, unlike the API method, is independent of the application. 
    Hence, it requires less modification to queries and code in the application layer since queries must be modified only to call UDFs.
    \item {\bf{The TDE Method:}} When you use Transparent Data Encryption (TDE), the encryption and decryption engines have to be installed right into the database engine.  
This encryption method takes place at the lowest system level and does not involve any modifications to the database environment or application’s source code. This means that administrators can easily install and manage the encryption engine in the database because there are no actions that need to be taken on the web server to do this.
 \end{itemize}
Database engines' ranking \cite{Ranking}  shows that APIs for encryption methods are commonly used for non-relational databases, whereas the most relational databases give three methods options.

\section{Access Control Mechanisms on Databases}

 \label{sec:ch4A.4}
Database engines' ranking \cite{Ranking} shows that the most common database engines employ Role-Based Access Control (RBAC) and passwords to validate access control rights.
Role-based access control (RBAC), also known as role-based security, is a system access control mechanism that assigns rights and privileges to give access to authorised users.
Furthermore, due to the increasing variety of roles in the real world, a rising number of (RBAC) roles are required to properly encapsulate the permissions, which in RBAC is an action or operation on an object or entity. This is known as "role explosion" \cite{elliott2010role}. An attribute-based approach to access control (ABAC) is often recommended, \cite{kuhn2010adding}. Role explosion does not feature in ABAC because there are no roles.

Attribute-based access control (ABAC) is an authorization mechanism that determines access based on attributes rather than roles. The goal of ABAC is to safeguard objects, such as data, against unauthorised access and interference. ABAC appeared as a kind of logical access control, developing from role-based access control (RBAC) and basic access control lists (ACL) that determine which users or system processes have access to objects and what actions are permitted with respect to certain objects. ABAC compares these component attributes with the rules. These rules say which combinations of attributes are allowed for the subject to do the right thing with the object.

This paper focuses on the security requirements of direct confidentiality, flexible access control, and querying over encrypted data 
at the data level. To fulfil the paper's aims, Ciphertext-Policy Attributes Based Encryption (CP-ABE) is discussed, which simultaneously provides data confidence (encryption) and access control at the same time. Conceptually, CP-ABE is comparable to traditional access control methods such as Role-Based Access Control (RBAC) \cite{bethencourt2007ciphertext}. In CP-ABE, the ciphertext is associated with an access policy, and the private key is associated with a set of attributes.  
However, in the case of querying over encrypted data, evidence is required, which is the focus of this section.

This paper employs a simplified version of the BSW scheme \cite{bethencourt2007ciphertext} that is sufficient for our purpose of assessing querying over encrypted data, while other versions have enhanced this version with additional features relating to complexity, hiding, and updating access control policies. Additionally, the BSW scheme is implemented in code in a variety of programming languages, including Java ( such as \cite{cp-abe14,cp-abe15}), C such as \cite{cp-abe7,cp-abe8}), C++ such as \cite{cp-abe12,cp-abe13}), and Python  such as \cite{cp-abe9,cp-abe10}), making it easy to utilise as required. On the other hand, enhanced CP-ABE schemes have a limited number of implementation, including Java ( such as \cite{cp-abe2B}) and JavaScript such as \cite{cp-abe1B}). In the next three sections, we'll talk about three things: CP-ABE functionality, CP-ABE security and performance analysis,  and how to evaluate CP-ABE in terms of speed and memory.

\subsection{CP-ABE Functionality on Database}
 \label{sec:4.5.2}
In this section, we will cover encrypted systems that use AES or CP-ABE to share data between one data owner and several users in order to facilitate flexible access and querying of encrypted data. CP-ABE is not supported by any database engines, hence AES and CP-ABE are assessed using the API method.

\subsubsection{Flexible Access control}
Many encryption systems, in practise, are natural examples of one-to-many communication, which refers to communication between the data owner and multiple users. 
A data owner's shared data with several users, the same key is used to encrypt and decrypt data   with symmetric encryption (i.e., in our case, AES). The key must be known by two or more communicating parties; no third party should have access to it. 
The paper in \cite{zhang2020attribute} explains how cloud computing access control works for new users. Here, how the data owner and users exchange data and keys and who can decrypt the data are covered. Therefore, we can determine the CP-ABE impact for both access control and the query.
When a new user joins the system, the data owner shares a secret key with this user. It's also important to set up another secret key when a user is revoked, and thus data owner can encrypt data with it again and share it with all other users who haven't been revoked.
This means that the owner should trust people who communicate with them, because getting this key puts all of the owner's data at risk.
Since shared secret keys are required for newly joined users, and re-encrypted data is required for revoked user events, this access control mechanism combined
with symmetric encryption is neither flexible nor scalable.
One-to-one public key cryptosystems can be used to protect one-to-many communication. These cryptosystems require that each message encrypted with a public key to be 
decrypted only with the associated secret key.
For example, a sender may encrypt data using a symmetric encryption key and then distribute this data key to all intended users through public key encryption. This technique is easy to implement but inefficient in terms of the number of encryption operations and the size of ciphertexts, which grows linearly with the number of users.
Another option in a one-way communication system is broadcast encryption\cite{boneh2005collusion}, which is meant to offer a mechanism for two unrelated parties to transmit a cryptographic key for data protection and other applications where a sender specifies a list of users or revoked users during encryption. Any intended user can decrypt with his secret key, but revoked users cannot even if they collaborate. Although broadcast encryption is efficient, it requires the description of each user separately. A sender must have a list of potential users on hand, as well as authorization information for each user. The person who owns the data or sends the message checks his user database with some important policy information to come up with a set of users for a specific message.
It is important in some system settings to be able to encrypt without knowing who the users who will be getting the information are.
Attribute-based encryption (ABE) provides the required ability to encrypt without knowing the specific settings of the users. Within the encryption process, it enforces access policies defined on attributes.
CP-ABE generates secret keys for each user based on the verification of the user's attributes.
When a new user joins,  they only need to set up a new secret key for the new user. Revocable CP-ABE is used to enable revocable access control when a user's access is revoked. In CP-ABE, there are two forms of revocation, which are direct revocation and indirect revocation. When a revocation event occurs, the nonrevoked users' secret keys should be updated since the revocation list is appended to the ciphertext in direct revocation. While in indirect revocation, nonrevoked users will be unaffected by revocation occurrences. The "entity revocation server" (RS) is updated with a "partially secret" key when a revocation event occurs.
These secret keys are distinct from the encryption key. Thus, if one of the keys is compromised, only a part of the data for this group, not the whole owner's data, is at risk. Fortunately, this access control mechanism in CP-ABE enables flexibility and scalability.

\subsubsection{Querying Over Encrypted Data}
The level of data leakage can be assessed by determining which computations 
(functionalities supported) over encrypted data can be executed. As for the AES functions, they are already defined as random (no computation) or deterministic (equality) operations as shown in the background, whereas CP-ABE has not been previously evaluated. 
The RaNDom (RND) and DETerministic (DET) operation via AES were used in the CryptDB\cite{popa2011cryptdb}  research to ensure  data confidentiality. CryptDB is a relational database system that enables queries to be executed on encrypted databases. Data is encrypted  by an onion that is built by layers. Each onion's layer supports a certain set of query types. When querying demands it based on types of computation, the required encryption layers may need to be removed.

CP-ABE uses symmetric encryption (i.e. AES) to encrypt data and access policies; hence, it is predicted that CP-ABE supports random (no computation) or/ and deterministic (equality) operations.
Therefore, both queries 1 and 2 were conducted  on plaintext and  data encrypted
with AES and CP-ABE. This experiment's objective is to identify the types of computation classes supported by CP-ABE and evaluate their impact on performance compared to AES. It is also intended to illustrate the possibility of applying it in an onion layer to provide data-level access control and confidentiality.
The experiment was conducted using a local MongoDB server and a client written in Java with their codes derived from \cite{wang2012java} for CP-ABE,\cite{Ajit2017java} for AES. It was conducted using a desktop PC with an Inter (R) Core (TM) 1.8 GHz processor and 8.00 GB, running Windows 10.
It is tested with 100 documents, with the following document fields: ($name$, $salary$, and $credit$ $card$ $number$). These queries were run $15$ times, and the average execution time was measured in milliseconds.
   \begin{verbatim}
    Q1: db.collection.find();
    Q2:db.collection.find(name: $eq "Alice");
\end{verbatim}

\begin{figure}[ht]
\centering
  \includegraphics[width=.5\textwidth,height=.18\paperwidth]{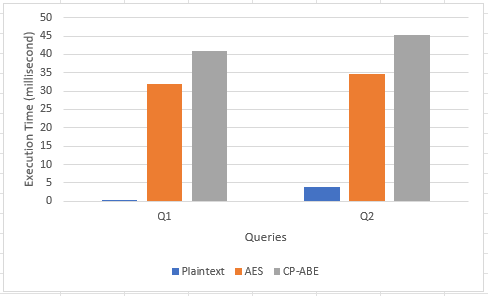}
  \caption{Result Analysis of using AES and CP-ABE}
  \label{fig:resultCP-ABE}
\end{figure}

\begin{table}[ht]

\centering

\begin{tabular}{|l|l|l|}
\hline
 & Q1 &  Q2\\ \hline
Plaintext & 0.47 & 3.93 \\ \hline
AES &31.86  & 34.72 \\ \hline
CP-ABE &40.93  & 45.41 \\ \hline
\end{tabular}
\caption{Average execution time of Q1 and Q2 (milliseconds)}
\label{tab:my-table}
\end{table}
As shown in Figure \ref{fig:resultCP-ABE}, a considerable time increase is observed in $Q1$ and $Q2$ in CP-ABE compared with AES. CP-ABE requires a longer execution time than AES, but it includes an extra access control security feature. Consequently, this performance trade-off increases security. 

\begin{figure}[ht]
\centering
  \includegraphics[width=.8\textwidth,height=.16\paperwidth]{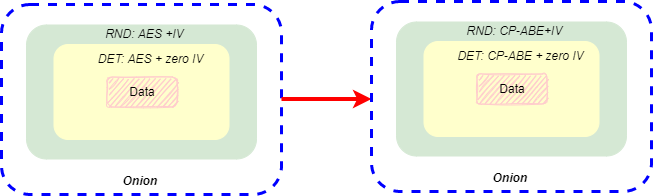}
  \caption{Replacement of AES with CP-ABE in Onion Layers.}
  \label{fig:AEStoCP-ABE}
\end{figure}

In short, this experiment reveals that CP-ABE exposes equal values as {\bf{\emph{Deterministic (DET)}}} Encryption. CP-ABE encrypts data and a padded access policy using AES with Zero IV. To prevent equality leakage and work as {\bf{\emph{Random (RND) }}}Encryption, CP-ABE can use a random IV rather than zero. 
As a result, these two layers could be replaced with CP-ABE as shown in Figure \ref{fig:AEStoCP-ABE}, which provides the same functionality but with less performance and increased security by restricting  who can access the data.

\subsection{CP-ABE and AES Security  and performance analysis}
 \label{sec:4.5.1}
This section analyses the security and performance of AES, which is a common algorithm  (see Section  \hyperref[sec:ch4A.3]{3}), comparing it to CP-ABE.
\begin{itemize}
\item {\bf{Security Analysis:}}
The use of CP-ABE and AES  encryption
is meant to keep data confidentiality and/or allow for flexible access control.
\begin{itemize}
\item {\emph{Data confidentiality:}} Cryptography was developed with the intention of 
preserving confidentiality by encryption.  If the data is confidential, only the intended receiver or recipients will be able to access or understand it. AES and CP-ABE ensure confidentiality.

\item {\emph{Flexible Access control:}} Each user has a unique CP-ABE decryption key that provides them with reading privileges for different parts of the data.
When all users have the same AES secret key to read and write, they can access all data encrypted with this secret key.
\end{itemize}
\item {\bf{Performance Analysis:}}
Each encryption algorithm has  its own  strengths and weaknesses that must be taken into consideration. Hence, it is necessary to understand the performance, strengths, and weaknesses of various cryptographic algorithms before deploying them to a particular application. As a consequence, these algorithms must be assessed in the context of multiple factors related to the evaluation efficiency of our proposal: (i) {\bf{Memory used}}, (ii) {\bf{Encryption Time}}, (iii) {\bf{Decryption Time}}. In this section, the following factors are considered for the comparison of encryption algorithms based on these factors.  The evaluation of encryption time and decryption time factors is performed using the CP-ABE algorithm ( the BSW \cite{bethencourt2007ciphertext})
in addition to AES with a CBC block.
\begin{itemize}
\item {\bf{Memory used :}}
The cost of a system is influenced by the amount of memory used. It is preferable if the amount of memory used is kept to a minimum. Different encryption algorithms require varying amounts of memory in their systems.
The size of the CP-ABE keys is determined by the complexity of the tree ($T$ ) and the number of attributes ($|w|$), whereas the size of AES keys remains constant. Moreover, the ciphertext size of CP-ABE is influenced by the attribute number of the access structure ($|A_c|$). The AES is based on the block size or padding length used. As a consequence, increasing the number of access tree attributes increases the ciphertext size for CP-ABE.
\item {\bf{Encryption time:}} 
It is the time it takes to convert plaintext to ciphertext. It has an impact on system costs. Thus, the timeframe should be kept as short as possible. Encryption time is influenced by the key size, the plaintext block size, and the encryption algorithm. The number of operations performed by the algorithm, such as the number of rounds on AES, the key size used, the number of initialization vectors used (e.g., $IV, r, p$), and the type of operations required by the method (e.g., "XOR" and "bilinear map"), all influenced computational complexity.Therefore, 
the complexity of the encryption of CP-ABE and AES, the length of text, the number of rounds, the type of block ciphers used, and the amount of padding used can all impact the AES encryption time, whereas the attribute number of the access structure influences the CP-ABE encryption time, wherein there is an increase in accordance with increases in the number of access tree attributes.
\item {\bf{Decryption time:}} It denotes the time required to retrieve plaintext from the ciphertext. The cost of a system is also determined by the time it takes to decrypt data. In the current case, this concerns efficient querying over encrypted data. In general, the decryption time should be less than the encryption time required to render the system more efficient. The variables that affect encryption time also impact the AES decryption time, while the number of user-assigned attributes to the private key has an influence on the CP-ABE decryption time. 

\end{itemize}

\end{itemize}
\subsection{{ Evaluation cost for CP-ABE and AES:}}
This section presents the results of our experiments, in addition to our analysis of the use of CP-ABE instead of AES. The memory used, the encryption time, and the decryption time performance are evaluated through encrypted files. First, the performance of the growth of CP-ABE attributes is measured, after which we assess the scalability performance with increasing data size related to encryption and decryption for both AES and CP-ABE with fixed attributes.
\\{\bf{Setup:}}
Experiments were performed using a desktop PC with an Inter Core i5 with  1.8 GHz processor and 8.00 GB, running Windows 10.

\begin{itemize}
 \item {\bf{\emph{Memory used:}}}The number of attributes $|A_C|$ in the access policy influences the ciphertext size; thus, it is first measured to see how it affects the overall size of the ciphertext in section \hyperref[sec:4.5.1]{4.2}. 
 CP-ABE with increasing the number of attributes, which raises the ciphertext's attribute set from 5 to 30, increasing by five each time, is used to measure ciphertext size on the disc drive  for 1000KB plaintext. 
Figure \ref{fig:StorageAtt} shows that increasing the number of attributes $|A_C|$ in the access policy linearly increases the size of the total ciphertext.
However, the size increase is slight related to the increase in attributes, since it was just 20 KB from 5 to 30 attributes.
Multiple files with increasing plaintext sizes (100KB, 200KB, 300KB, 400KB, 500KB, 600KB, 700KB, 800KB, 900KB, and 1000KB) with an increasing  number of attributes are encrypted by CP-ABE, and the  ciphertext size is second measured by the rate of increase compared to the size of the plaintext.
Figure  \ref{fig:StorageAttCPABE2} shows that the CP-ABE ciphertext is twice as large as the plaintext and attributes have little effect on overall size.
Figure \ref{fig:DZCPABE} compares AES and CP-ABE for multiple files of increasing size with the minimum and maximum attributes in the above experiment. It demonstrates that the ciphertext size of both CP-ABE and AES increases linearly with plaintext size, while AES requires less space than  CP-ABE regardless of the number of attributes.
\begin{figure}[ht]
     \centering
     \hfill
     \begin{subfigure}[b]{0.48\textwidth}
         \centering
  \includegraphics[width=\textwidth,height=.25\paperwidth]{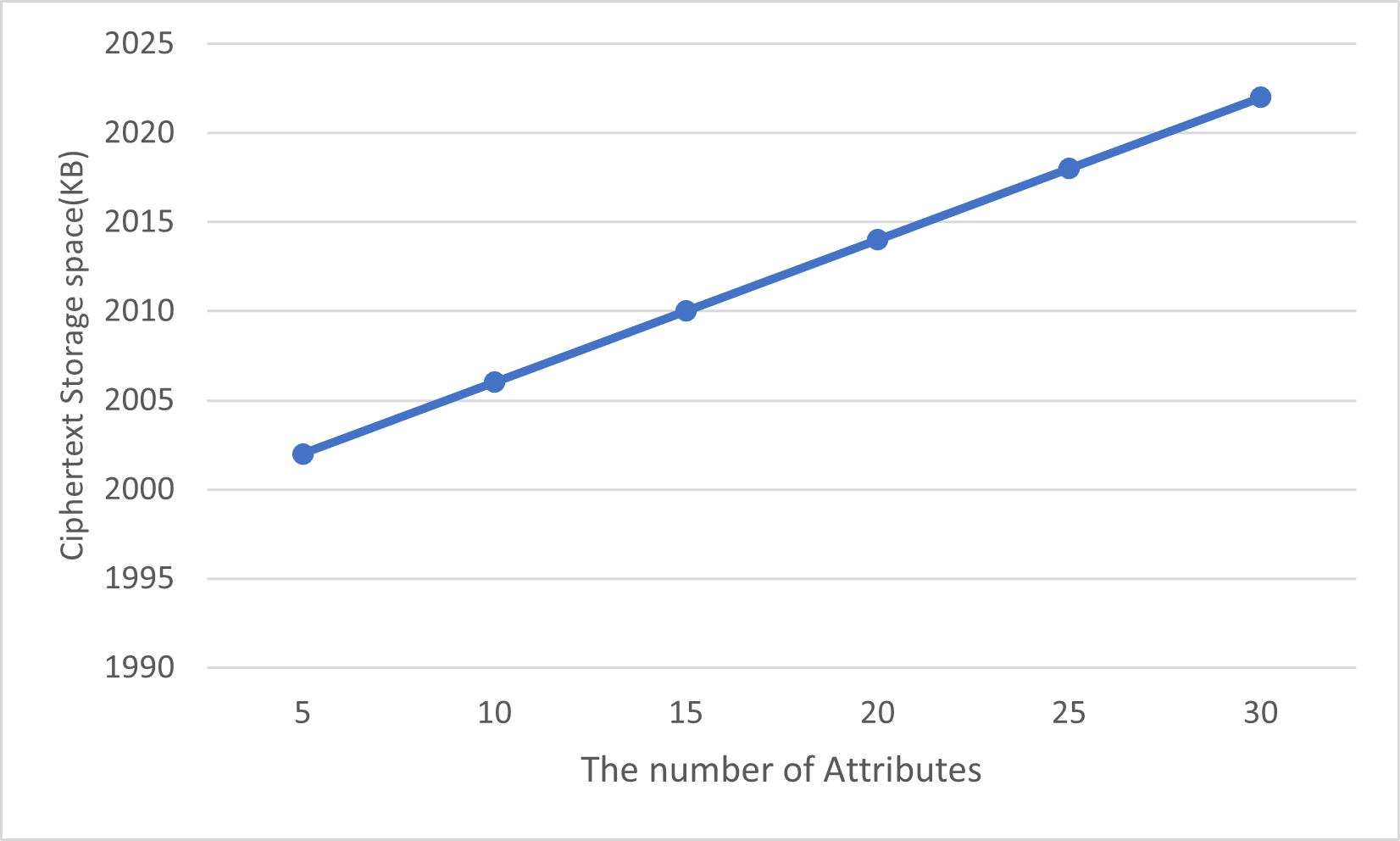}
\caption{CP-ABE Ciphertext size with growing attributes}
\label{fig:StorageAtt}
     \end{subfigure}
     \hfill
     \begin{subfigure}[b]{0.48\textwidth}
         \centering
       \includegraphics[width=\textwidth,height=.25\paperwidth]{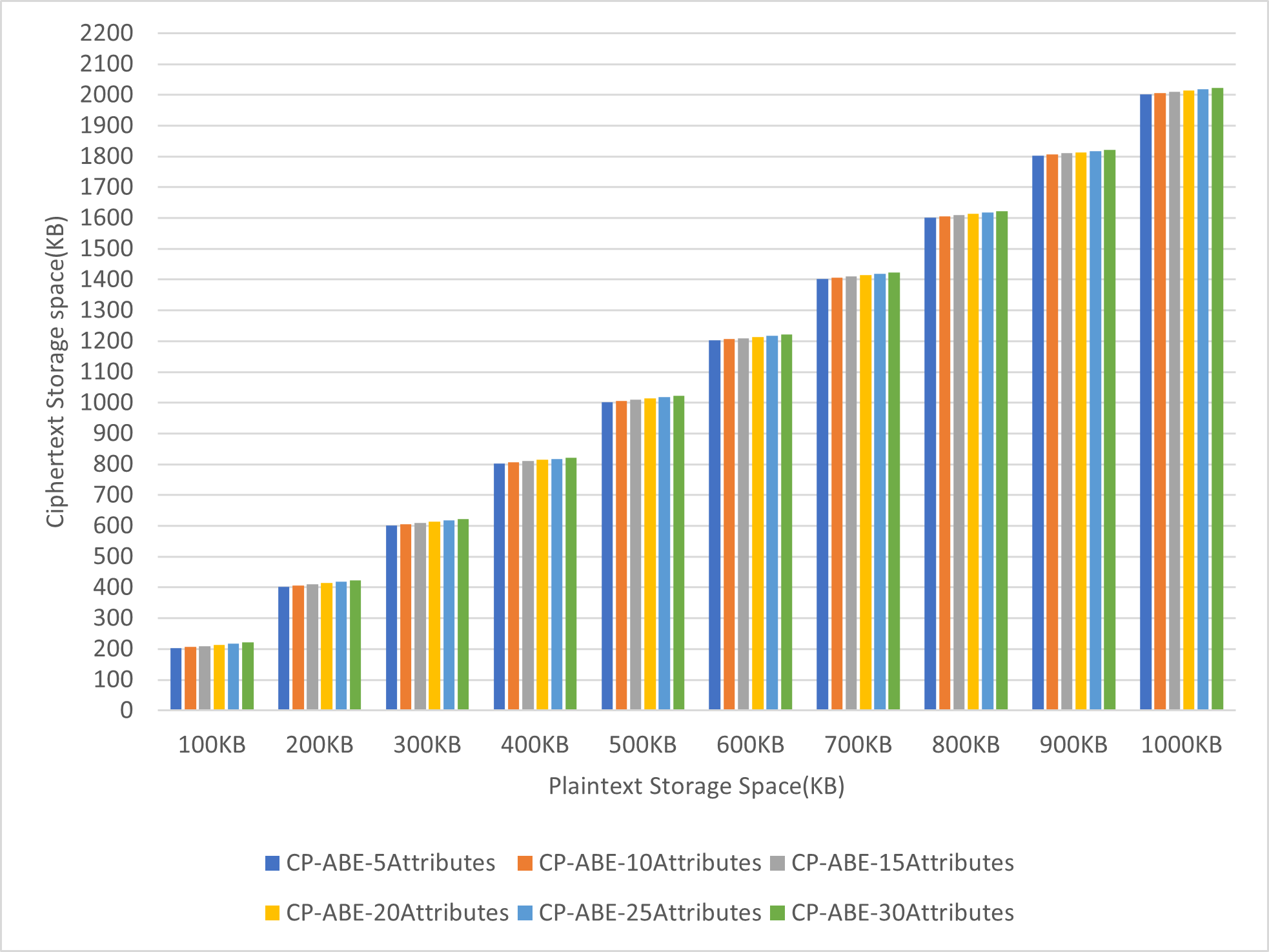}
\caption{CP-ABE ciphertext  with  growing Attributes}
\label{fig:StorageAttCPABE2}
     \end{subfigure}
     \hfill
     \begin{subfigure}[b]{0.48\textwidth}
         \centering
       \includegraphics[width=\textwidth,height=.25\paperwidth]{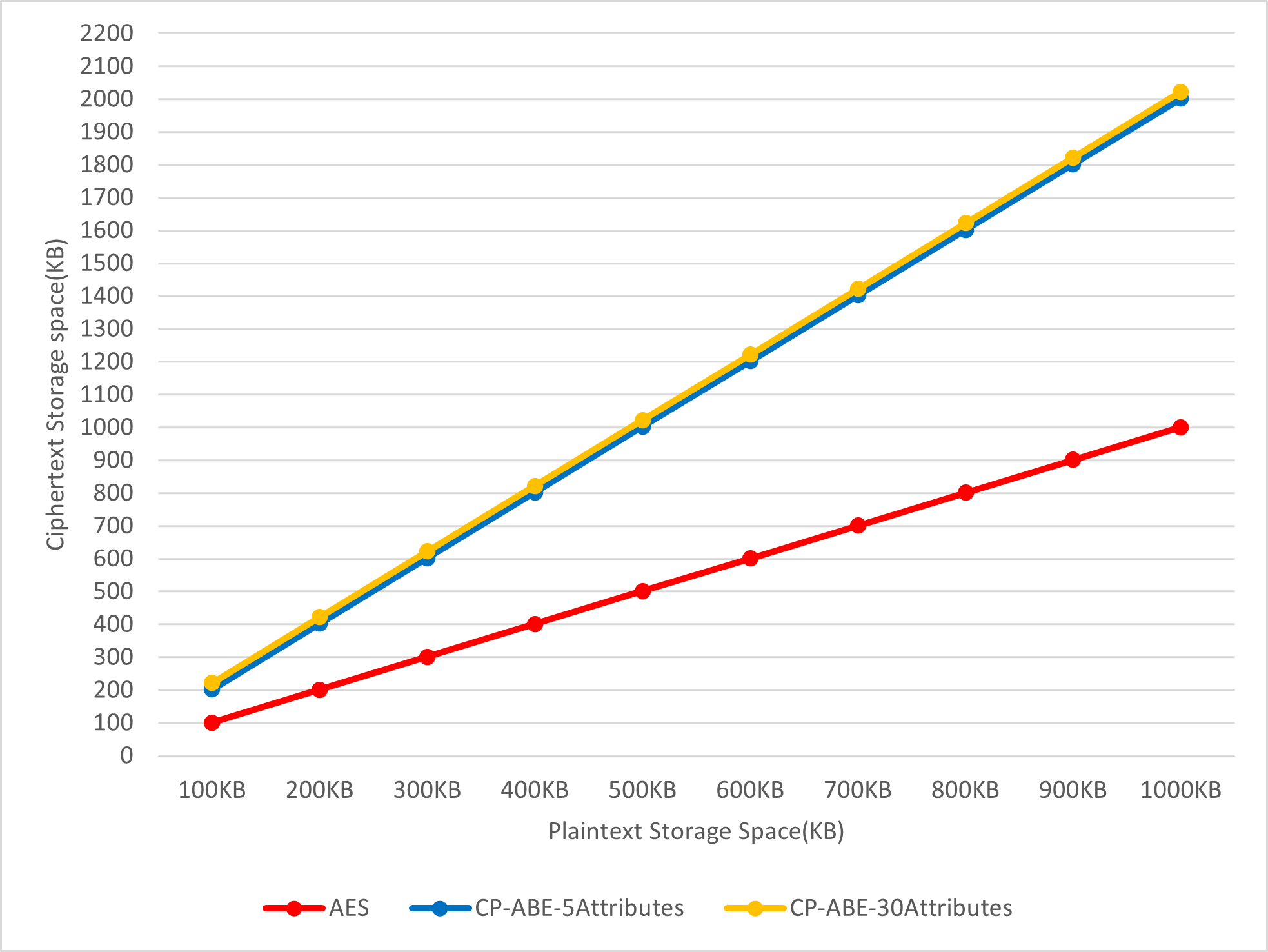}
\caption{ Comparison of  ciphertext size for CP-ABE vs. AES}
\label{fig:DZCPABE}
     \end{subfigure}

\caption{Memory used of CP-ABE}

\end{figure}
\item {\bf{\emph{Encryption Time: }} }It takes more time to encrypt a file if there are more attributes in the access policy.
In the first part, we examine the CP-ABE encryption costs with the number of attributes growing. A file with a constant size of 100KB, 200KB, 300KB, 400KB, 500KB, 600KB, 700KB, 800KB, 900KB, and 1000KB is used with an increasing number of attributes in the ciphertext, wherein the ciphertext attribute set ranges from 5 to 30, increasing by five each time. 
Figure \ref{fig:EXEncAttCPABEA} depicts the time cost of encryption, which increases significantly as the number of attributes increases for a constant data size.
Second, for both CP-ABE with a fixed number of attributes (5 and 30) and AES, the total execution encryption is calculated in accordance with multiple files with increasing plaintext sizes that are encrypted. The encryption time is measured.
Figure  \ref{fig:EXEncDCPABE2} demonstrates that both CP-ABE and AES can increase the encryption time linearly with plaintext size. However,   rising CP-ABE is also linked to a number of attributes ($|A_C|$) but AES takes less time to execute than CP-ABE regardless of the number of attributes. 
\begin{figure}[ht]
     \centering
     \hfill
     \begin{subfigure}[b]{0.48\textwidth}
         \centering
  \includegraphics[width=\textwidth,height=.25\paperwidth]{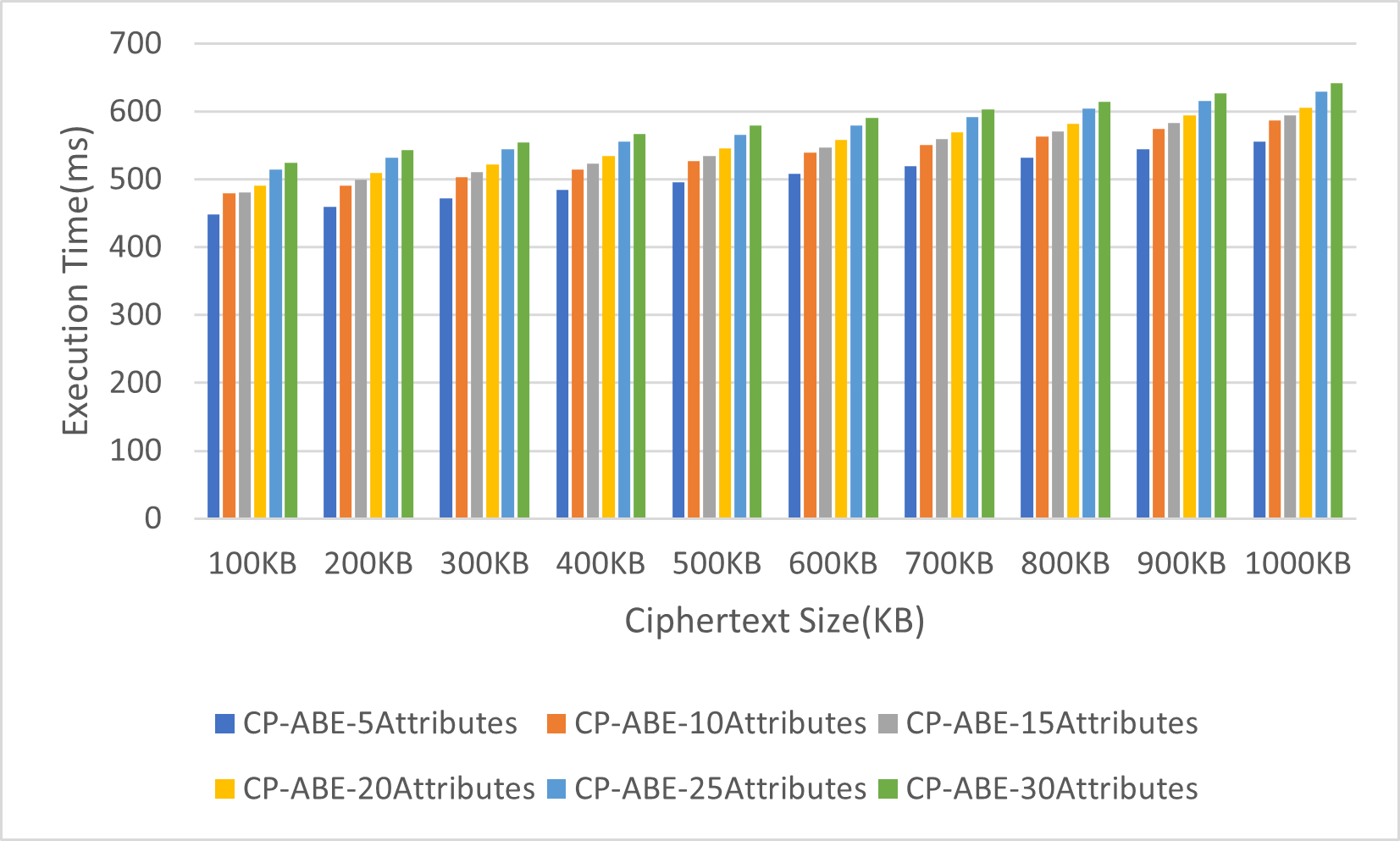}
\caption{CP-ABE Encryption Time with  growing Attributes }
\label{fig:EXEncAttCPABEA}
     \end{subfigure}
     \hfill
     \begin{subfigure}[b]{0.48\textwidth}
         \centering
       \includegraphics[width=\textwidth,height=.25\paperwidth]{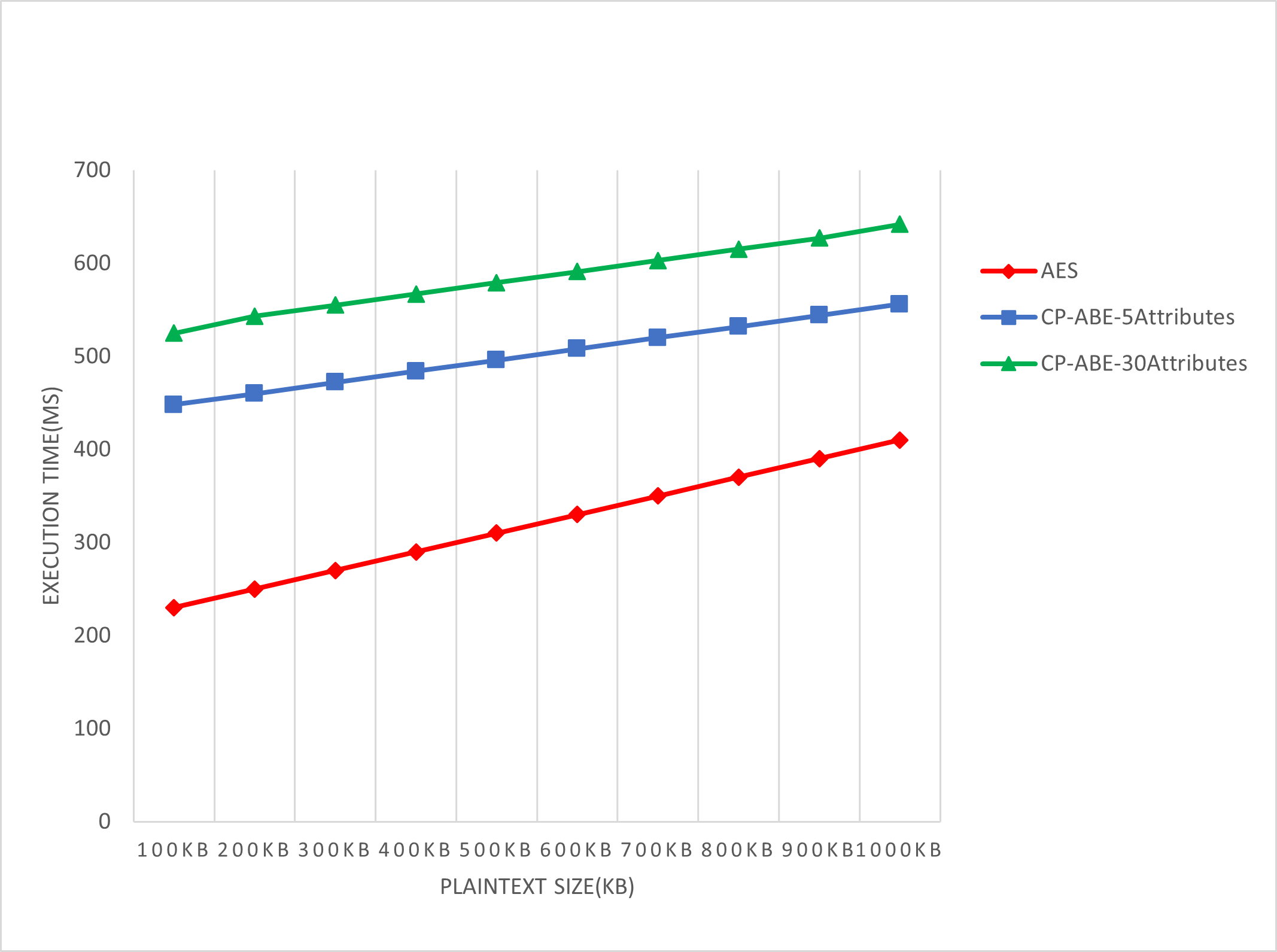}
\caption{ Comparison of Encryption Time for  CP-ABE Vs. AES  }
\label{fig:EXEncDCPABE2}
     \end{subfigure}
     \hfill

\caption{Encryption execution Time of CP-ABE}

\end{figure}

\item {\bf{\emph{Decryption Time: }} } The number of attributes $|S|$ in the private key has also an effect on decryption time.
In the first part, we examine the CP-ABE decryption costs with the number of attributes in the private key growing. A file with a constant size of  100KB, 200KB, 300KB, 400KB, 500KB, 600KB, 700KB, 800KB, 900KB, and 1000KB is used with an increasing number of attributes in a private key, wherein the attribute set ranges from 5 to 30, increasing by five each time.  The attribute set is included in the decryption cost. The time cost of decryption increases as the number of attributes grows in complexity, as in Figure \ref{fig:EXdecAttCPABE2}. Second, the decryption time is measured with growing plaintext size. Figure  \ref{fig:EXDecDCPABE} reveals that both CP-ABE and AES increase decryption time in an almost linear ways with the plaintext size, whereas AES's decryption time takes less time to execute than CP-ABE and  decryption time is faster than encryption timefor the same plaintext size for both. Rising CP-ABE is also associated with the number of attributes $|S|$ in the private key.

\begin{figure}[ht]
     \centering
     \hfill
     \begin{subfigure}[b]{0.48\textwidth}
         \centering
  \includegraphics[width=\textwidth,height=.25\paperwidth]{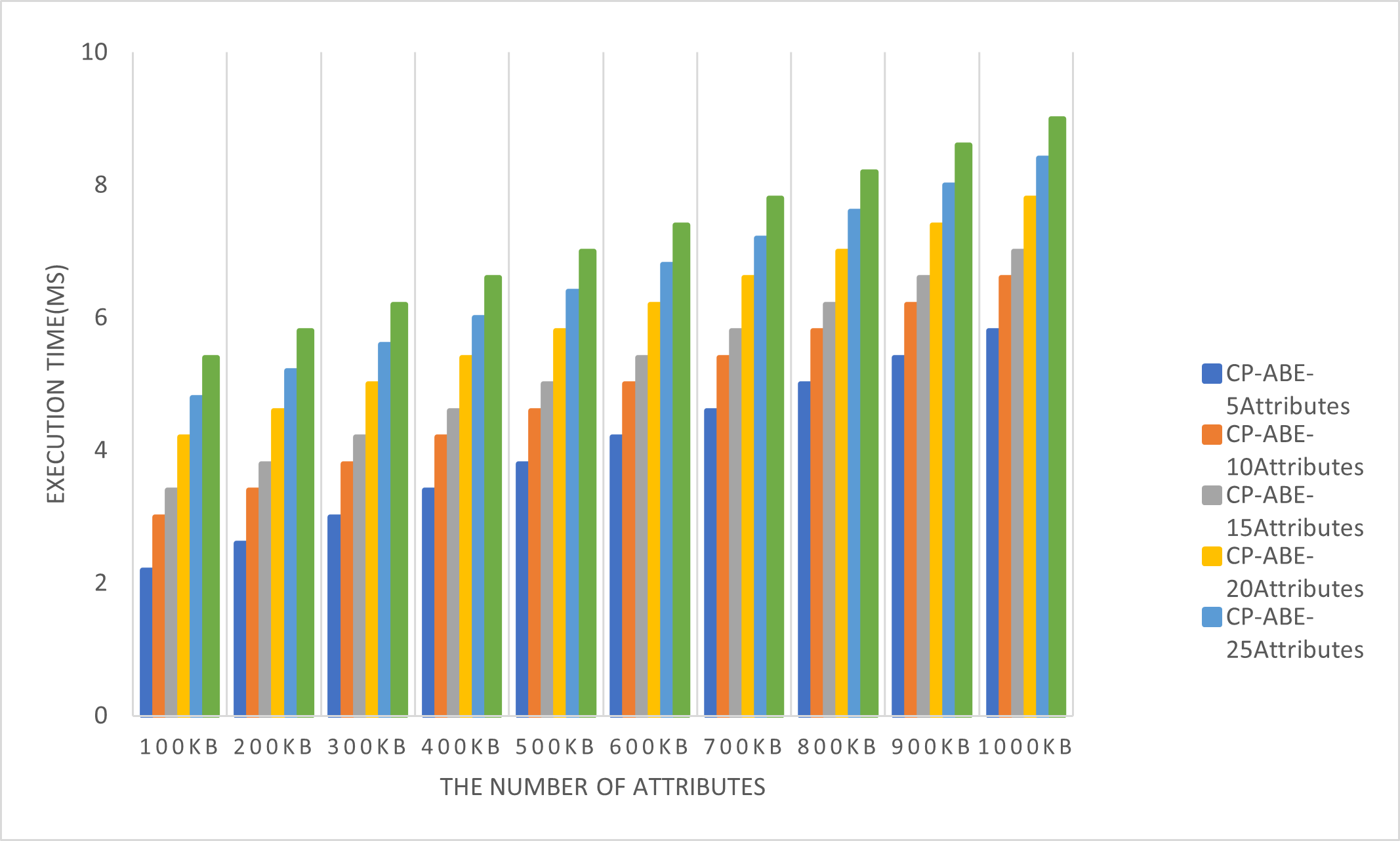}
\caption{ Comparison of execution Time for  CP-ABE Vs. AES   }
\label{fig:EXdecAttCPABE2}
     \end{subfigure}
     \hfill
     \begin{subfigure}[b]{0.48\textwidth}
         \centering
       \includegraphics[width=\textwidth,height=.25\paperwidth]{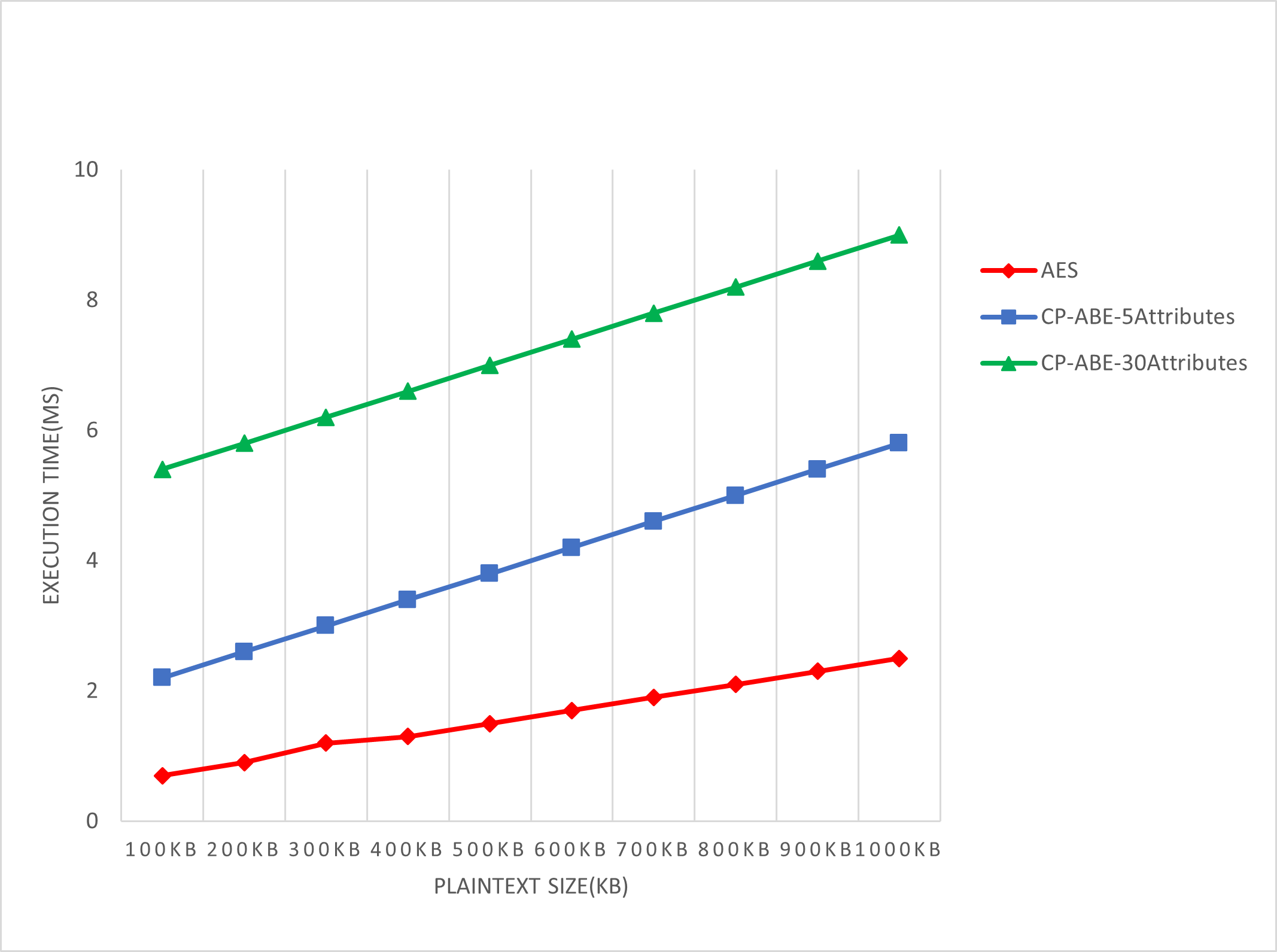}
\caption{Execution Time of CP-ABE and AES with  growing data size }
\label{fig:EXDecDCPABE}
     \end{subfigure}
     \hfill

\caption{Decryption execution Time of CP-ABE}

\end{figure}

\end{itemize}

\section{Conclusion}
 \label{sec:ch4A.6}

The purpose of this paper is to determine the computation classes supported by CP-ABE and assess their impact on performance. It is also intended to demonstrate how it could be used in an onion layer to provide data-level access control and confidentiality.
CP-ABE is often used to transfer files between the owner and the user (s). However, it was not previously addressed if the data owner desires to store CP-ABE-encrypted data in a database and then allow users to query this data in accordance with the owner's policies.
Therefore, this paper examines the fundamental concept of Ciphertext Policy Attribute-Based Encryption, which simultaneously helps to  satisfy three major security database requirements, namely: confidentiality, querying over encrypted data, and flexible access control at the data level.
 Furthermore, ciphertext policy attribute-based encryption and its impact were suggested instead of AES and evaluated. CP-ABE decreases performance in comparison to AES,  when the number of attributes in the access policy  and data size are increased. However, it provides more security. CP-ABE exposes equal values since it encrypts data and has a padded access policy using AES with Zero IV. To avoid equality leaking, CP-ABE can be updated using a random IV rather than a zero-IV.

\bibliographystyle{unsrt}  
\bibliography{references}

\end{document}